\documentclass[apjl,twocolumn]{openjournal}
\usepackage{amsmath}
\usepackage{booktabs}
\usepackage{multirow}
\usepackage{color}
\usepackage{soul}
\usepackage{upgreek}
\usepackage{natbib}
\usepackage[normalem]{ulem}
\usepackage{comment}

\usepackage[dvipsnames]{xcolor} 

\usepackage{hyperref}
\hypersetup{
    colorlinks=true,
    linkcolor=blue,
    breaklinks=true,      
    urlcolor=blue,
    citecolor=blue,
}
\usepackage{orcidlink}

\renewcommand{\arraystretch}{1.2}  
\usepackage{listings}
\usepackage{color}
\definecolor{dkgreen}{rgb}{0,0.6,0}
\definecolor{gray}{rgb}{0.5,0.5,0.5}
\definecolor{mauve}{rgb}{0.58,0,0.82}
\definecolor{golden}{rgb}{0.86,0.65,0.01}
\begin{document}

\title{The abundance and properties of the lowest luminosity dwarf galaxies around the Milky Way: Insights from Semi-Analytic Models \vspace{-1.5cm}}

\author{
Niusha Ahvazi,$^{1,2,3 \star \dagger} \orcidlink{0009-0002-1233-2013}$%
Andrew B. Pace,$^{1 \dagger} \orcidlink{0000-0002-6021-8760}$
Christopher T. Garling, $^{1} \orcidlink{0000-0001-9061-1697}$
Xiaowei Ou, $^{1,2,3 \dagger} \orcidlink{0000-0002-4669-9967}$
Nitya Kallivayalil, $^{1,3} \orcidlink{0000-0002-3204-1742}$
Paul Torrey, $^{1,2,3} \orcidlink{0000-0002-5653-0786}$
Andrew Benson,$^{4} \orcidlink{0000-0001-5501-6008}$
Aklant Bhowmick,$^{1,2,3 \dagger} \orcidlink{0000-0002-7080-2864}$
Núria Torres-Albà, $^{1 \dagger} \orcidlink{0000-0003-3638-8943}$
Alex M. Garcia, $^{1,2,3} \orcidlink{0000-0002-8111-9884}$
Alejandro Saravia, $^{1} \orcidlink{0000-0003-4546-3810}$
Jonathan Kho, $^{1,2,3} $
Jack T. Warfield, $^{1} \orcidlink{0000-0003-1634-4644}$
Kaia R. Atzberger $^{1} \orcidlink{0000-0001-9649-8103}$ 
\\
$^{1}$Department of Astronomy, University of Virginia, 530 McCormick Road, Charlottesville, VA 22904, USA\\
$^{2}$ Virginia Institute for Theoretical Astronomy, University of Virginia, Charlottesville, VA 22904, USA\\
$^{3}$ The NSF-Simons AI Institute for Cosmic Origins, USA\\
$^{4}$Carnegie Observatories, 813 Santa Barbara Street, Pasadena, CA 91101, USA\\
}

\thanks{$^\star$E-mail:} 
\email{nahvazi@virginia.edu}
\thanks{$^\dagger$ Galaxy Evolution and Cosmology (GECO) Fellow}

\begin{abstract}
We investigate the formation and observable properties of faint satellite galaxies ($M_V > -3$) in Milky Way–like halos using the semi-analytic galaxy formation model {\sc Galacticus}. The ability of the smallest dark matter halos to form stars depends sensitively on the balance between gas cooling and reionization heating. To quantify how this balance shapes the abundance and properties of the faintest galaxies, we compare two model variants: a fiducial model that includes molecular hydrogen (H$_2$) cooling and UV background radiation, and a No-H$_2$ model with atomic cooling only. Both models reproduce the structural properties of brighter Milky Way satellites, but they diverge at the lowest luminosities in the hyper-faint regime. The fiducial model predicts a substantially larger population of such systems that are on average hosted in halos with lower peak masses and quenched earlier. Many of these predicted systems lie below current observational thresholds but are within reach of next-generation deep imaging surveys. The predicted size–luminosity distributions of both models overlap with the region occupied by recently discovered “ambiguous” systems, whose classification as galaxies or star clusters remains uncertain. Specifically, we find that hyper-faint satellites have line-of-sight velocity dispersions of $\sigma_{\rm los} \sim 1–3$ km/s in the fiducial model, nearly an order of magnitude higher than expected for purely self-gravitating stellar systems of the same stellar mass. This distinction underscores the diagnostic power of precise kinematic measurements for determining whether ambiguous objects are dark matter dominated dwarf galaxies or star clusters, and highlights the importance of upcoming spectroscopic campaigns in resolving the nature of the faintest satellites.


\end{abstract}

\maketitle  

\section{Introduction}

Dwarf galaxies are powerful probes of galaxy formation physics and the nature of dark matter. Their abundance, internal structure, and star-formation histories reflect the interplay between baryonic processes and the underlying dark-matter potential \citep[e.g.,][]{Wetzel2016, Bullock2017, Sales2022}. In the Local Group, satellite galaxies provide one of the observational windows into galaxy formation at halo masses below 10$^9$ M$_\odot$, a regime where physical processes such as cooling \citep{Nadler2025}, feedback \citep{Read2016, Tollet2016, Fitts2017}, and environmental effects \citep[e.g.,][]{Font2008, Simpson2018} operate in concert to determine whether a halo hosts a luminous/visible galaxy.

Over the past decade, wide-area photometric surveys and Gaia astrometry have revealed a growing population of extremely low-luminosity stellar systems around the Milky Way (MW; e.g., SDSS: \citealt{Willman2005,Belokurov2006}; PS1: \citealt{Laevens2015,Laevens2015b}; DES: \citealt{Bechtol2015,Drlica-Wagner2015}; DELVE: \citealt{Cerny2021}; HSC-SSP: \citealt{Homma2016,Homma2018}), often referred to as ultra-faint satellites \citep[e.g., M$_V>-7.7$][]{Simon2019}. Many recently discovered systems lie at the boundary between dwarf galaxies and compact stellar systems, complicating their classification. Catalogs and compilations \citep[e.g.,][]{Pace2024} now include a number of such “ambiguous” objects whose structural and kinematic properties place them near the transition between dark matter–dominated dwarfs and dark matter–free star clusters. In this work, we adopt the term {\it hyper-faint} satellites to refer specifically to the subset of these systems that would be classified as bona fide satellites if they are confirmed to reside in dark-matter halos. Quantitatively, we define hyper-faints as systems with absolute $V$-band magnitudes M$_V > -3$ and projected half-light radii $1 \lesssim R_h~[\mathrm{pc}] \lesssim 10$, corresponding to the extreme small-size tail of the broader ultra-faint population.

Understanding whether these ambiguous systems are bona fide galaxies or stellar clusters has important consequences for our galaxy formation models. If they are dark-matter–dominated, they probe galaxy formation in the smallest halos and place constraints on the low-mass end of the halo mass function and on dark matter models \citep[e.g.,][]{Nadler2021}. If they are star clusters, they instead inform us about the formation and survival of bound stellar systems and the limits of cluster formation.

At a physical level, the fundamental distinction between dwarf galaxies and star clusters in the $\Lambda$CDM framework is the presence (or absence) of a dark matter halo. Dwarf galaxies are dark matter–dominated systems \citep{Willman2012}, while star clusters are believed to be purely baryonic. Observationally, this difference manifests in a set of structural, kinematic, and chemical diagnostics. Dwarf galaxies generally exhibit larger half-light radii/sizes at fixed luminosity, larger velocity dispersions indicative of dark matter, and spreads in stellar metallicity and element-abundance ratios---signatures of extended or multi-epoch star formation and self-enrichment {\citep[e.g.,][]{Willman2012,Walker2009,Kirby2008}. Globular clusters, by contrast, are typically compact, and have velocity dispersions consistent with purely stellar systems. While some globular clusters show abundance anomalies and evidence for multiple stellar populations, the associated metallicity spreads are generally small compared to those found in dwarf galaxies \citep{Carretta2009, Roediger2014}.

Recently discovered/classified ambiguous systems occupy the region of size–luminosity parameter space where these diagnostics overlap. For many of these objects only sparse spectroscopic data exist, often yielding upper limits on line-of-sight velocity dispersion or small samples prone to contamination and binary-star biases (see discussion in \citealt{Simon2019}). Chemical abundance information, such as measurable [Fe/H] spreads \citep{Kirby2011, Kirby2013}, has emerged as a complementary discriminator, with dwarf galaxies generally exhibiting larger metallicity spreads due to multi-epoch star formation, whereas stellar systems tend to be chemically homogeneous.

From the observational side, incomplete survey coverage complicates comparisons between theory and data. Surface-brightness limits, variable depth across surveys, and detection biases strongly affect the census of low-surface-brightness and small-size systems; careful modeling of survey selection functions is required to transform theoretical predictions into observable quantities (e.g., methods and completeness estimates in \citealt{Drlica-Wagner2020}; projected LSST sensitivity estimates in \citealt{Tsiane2025}). Additionally, foreground contamination by MW stars, membership uncertainties, and unresolved binaries introduce systematics into property estimations (e.g. velocity-dispersion and metallicity measurements).

On the theoretical side, modeling galaxy formation and evolution of satellite galaxies in the lowest-mass halos is particularly challenging because it requires simulations that can resolve a large dynamic range, tracking subhalos several orders of magnitude smaller than their host while still modeling the larger-scale environment. This demands accurate tracking of satellite orbits within their host halos and the ability to follow tidally stripped subhalos without artificially disrupting them \citep{vandenBosch2018,vandenBosch2018b}. Environmental processes such as tidal stripping, tidal heating, and ram-pressure stripping can further alter the structural and dynamical properties of satellites over time.

In addition to environmental effects, key internal and external physical processes—such as cooling, star-formation efficiency, stellar feedback, and the effect of reionization—play an important role in setting the stellar mass content of the smallest subhalos. Among these, the interplay between gas cooling and reionization is particularly crucial for determining whether a halo can form stars at all. Including molecular hydrogen (H$_2$) cooling enhances the efficiency of gas cooling and star formation in halos with virial temperatures below $10^4$ K \citep[e.g.,][]{Abel1995, Tegmark1997}. However, the ultraviolet (UV) background radiation generated during reionization can photodissociate H$_2$ and heat the intergalactic medium (IGM), suppressing gas accretion into the shallow potential wells of low-mass halos \citep[e.g.,][]{Ocvirk2016}. The competition between these processes ultimately determines the minimum halo mass capable of hosting a luminous galaxy. Understanding this balance is therefore essential for predicting the abundance and observable properties of the faintest satellites.

Given these challenges, semi-analytic models (SAMs) offer an efficient approach to explore how faint satellite populations depend on uncertain baryonic physics. In this work, we use the {\sc Galacticus} SAM (\citealt{Benson2012Galacticus}), configured to probe H$_2$ chemistry and UV background models, to (i) predict the abundance and properties of hyper-faint satellites around MW analogs, and (ii) assess the extent to which H$_2$ cooling alters the threshold for galaxy formation at the lowest masses, and quantify the role of H$_2$ cooling in setting the galaxy formation threshold. We compare these model predictions with current observational compilations of MW satellites, including ambiguous systems cataloged in \cite{Pace2024}, taking survey incompleteness into account following detection probability models for DES, PS1, and LSST by \cite{Drlica-Wagner2020, Tsiane2025}. We evaluate the diagnostic power of kinematic measurements (line-of-sight velocity dispersion) in determining the nature of ambiguous systems.

This paper is organized as follows. In Section~\ref{method}, we describe the SAM and its implementation. Section~\ref{result} presents our results, while Section~\ref{discussion} discusses their implications in the context of observations and modeling caveats. Finally, Section~\ref{conclusions} summarizes our findings and highlights key conclusions.

\section{Method}\label{method}

We use the {\sc Galacticus} semi-analytic model (SAM) of galaxy formation and evolution \citep{Benson2012Galacticus}\footnote{The specific version used in this work is publicly available at \href{https://github.com/galacticusorg/galacticus/commit/ee197e314ac28f1eca3b6dcc47c5f88682bbddb5}{https://github.com/galacticusorg/galacticus}}, which solves coupled differential equations to model the key physical processes that govern galaxy formation over cosmic time, evolving galaxies along the branches of dark matter halo merger trees. For a detailed description of the model and its configuration, see \cite{Ahvazi2024}.

In brief, we construct halo merger trees using the extended Press-Schechter formalism \citep{Press1974,Parkinson2007} with the modified merger rates from \citet{Benson2017}. This approach allows us to generate large statistical samples of MW-like halos while resolving substructures down to $10^7 \,\mathrm{M}_\odot$, sufficient to model the formation of luminous low-mass satellite galaxies.

The physics in {\sc Galacticus} includes prescriptions for gas cooling, star formation, stellar feedback, gas re-incorporation, galaxy merging, and subhalo orbital evolution (including dynamical friction and tidal stripping). Gas cooling is modeled following the formalism of \citet{White-Frenk1991}, where the cooling radius is determined as the radius within a dark matter halo where the local gas cooling time equals the dynamical timescale of the halo. Gas inside this radius is able to cool and accrete onto the galaxy, while gas outside remains hot and in hydrostatic equilibrium. This framework naturally distinguishes between rapid “cold” accretion, when the cooling radius exceeds the virial radius, and quasi-static cooling, when it lies within the halo}. Metallicity-dependent atomic hydrogen cooling rates are calculated using CLOUDY (v23.01; \citealt{Gunasekera2023}) under the assumption of collisional ionization equilibrium. Star formation in the disk is modeled using a hydrostatic pressure-based formulation by \citet{Blitz2006}, and for the spheroidal component star formation rates depend on the gas mass content and dynamical time of the system. Feedback is implemented using a power-law prescription that relates the outflow rate to the energy input from stellar populations, with separate treatments for disk and spheroid components. Ejected gas is stored in an outflow reservoir and gradually reincorporated into the hot halo on a timescale proportional to the halo dynamical time. Summaries of these implementations can be found in \citet[Section 2.2]{Knebe2018}, \citet[Section 2.1]{Weerasooriya2023}, and \citet[Appendix A]{Ahvazi2024}. 

This study focuses on the role of molecular hydrogen (H$_2$) contributions to the circumgalactic medium (CGM) cooling, particularly its impact on the threshold for galaxy formation and the abundance and properties of very low-mass MW satellites. To isolate this effect, we compare two models; \textbf{No-H$_2$ Model}: a baseline model in which gas cooling proceeds exclusively via atomic hydrogen. The atomic cooling rates are computed following \citet[][see their Figure 3]{Benson2012Galacticus} for collisional ionization equilibrium with metallicity-dependence to determine the cooling function\footnote{Note that in this work we use the updated version of CLOUDY v23.01, whereas \citealt{Benson2012Galacticus} employed the older v08.00 release}. \textbf{Fiducial Model}: identical to the No-H$_2$ model but with the addition of H$_2$ cooling and associated physics.

In the fiducial model we track H$_2$ formation and destruction using the chemical network from \citealt{Abel1997} (for more details see \citealt[Section 2.3]{Ahvazi2024}), and assume the CGM temperature equals the virial temperature of its host halo. The CGM is treated as a uniform-density sphere containing the current CGM mass, with its radius set to the virial radius for isolated halos and the ram-pressure stripping radius for satellites (modeled following \citealt{Font2008}). To account for CGM density enhancements in the inner regions, we apply a clumping factor (see Eq. 2 in \citealt{Ahvazi2024}).
Photodissociation of H$_2$ due to the UV background radiation is also included. We adopt the \citet{FaucherGiguere2020} model to compute the cosmic background radiation field, interpolating the spectral radiance as a function of redshift and wavelength. Self-shielding of H$_2$ against this background is implemented using the prescription from \citet[Eq.~11]{SafranekShrader2012}. With the resulting H$_2$ abundance, we compute its contribution to the cooling function, $\Lambda(T)$, using the fitting functions of \citet{Galli1998}.

For each of the two cooling prescriptions, we model a suite of 100 MW analogs, each with a $z = 0$ virial mass of $10^{12} \, \mathrm{M}_\odot$. This sample size allows us to capture halo-to-halo variations in satellite populations arising from differences in assembly histories. The resulting satellite properties are compared against MW satellite observations compiled by \citet{Pace2024}\footnote{For access to the dataset, see the accompanying GitHub repository at \href{https://github.com/apace7/local_volume_database}{https://github.com/apace7/local\_volume\_database}}.

\begin{figure*}
	\includegraphics[width=2.09\columnwidth]{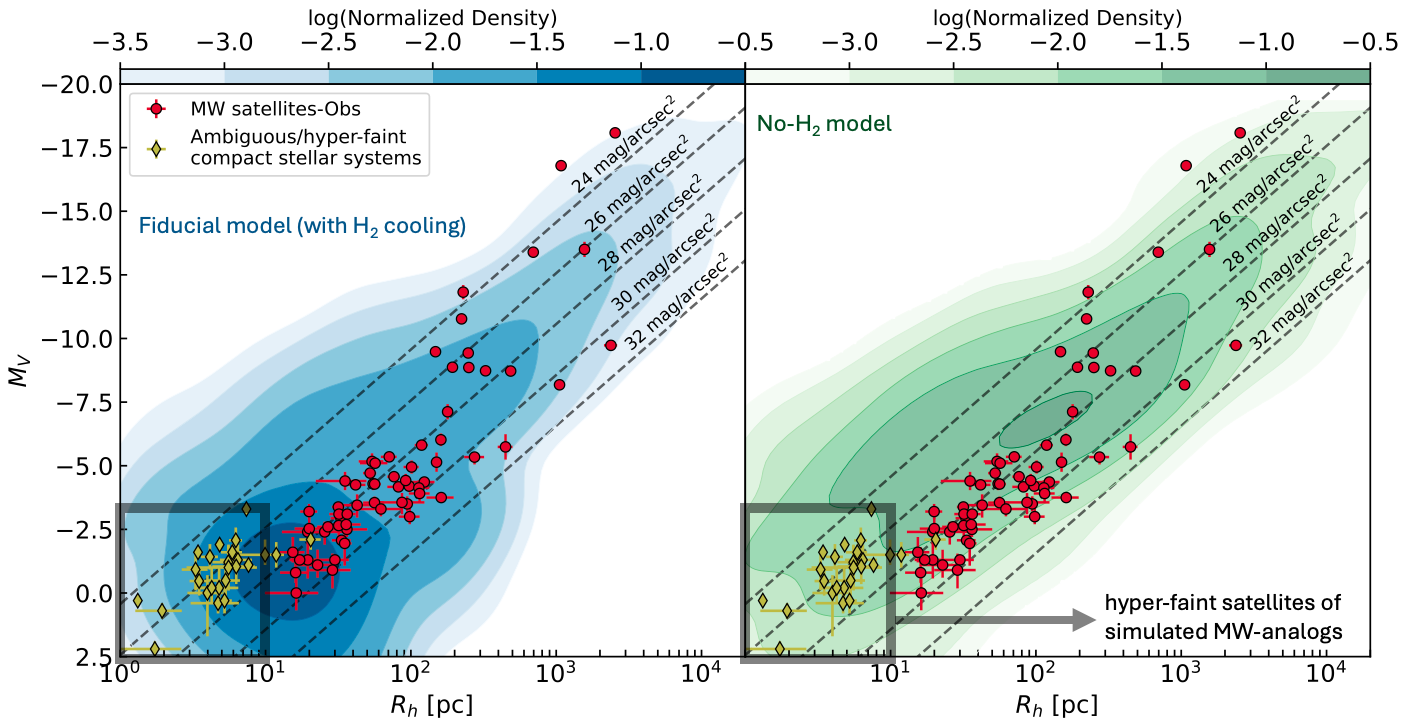}
    \caption{Absolute V-band magnitude (M$\rm_{V}$) as a function of projected half mass radius (R$\rm_{h}$) predictions from our model for satellites of MW-analogs. The left panel shows results from the fiducial model (including H$_2$ cooling), while the right panel shows the No-H$_2$ model (atomic cooling only). Model predictions are shown as shaded regions, where the color scale indicates the logarithm of the normalized probability density of satellites (see color bar at the top of each panel). Dashed diagonal lines correspond to constant surface brightness. Confirmed MW satellites (red circles) and ambiguous objects (olive diamonds) are shown for comparison. The gray box highlights the region associated with hyper-faint systems ($M{\rm _V} > -3$ and $1 \lesssim R{\rm _h} [pc] \lesssim 10$) on the size–luminosity relation.}
    \label{fig:size_luminosity}
\end{figure*}

\subsection{Observational incompleteness}
\label{sec:obs_incompleteness}

When comparing our model predictions with observations of MW satellite galaxies, it is essential to account for observational incompleteness. This is particularly important when comparing satellite luminosity functions, as the detectability of faint and diffuse systems is limited by survey depth, resolution, and surface brightness sensitivity. Detection probability estimates from the DES and PS1 surveys \citep{Drlica-Wagner2020}, LSST \citep{Tsiane2025}, and the DELVE survey \citep{Tan2025} provide frameworks for correcting model predictions for incompleteness.

In \citet{Drlica-Wagner2020} and \citet{Tsiane2025}, detection efficiency is quantified by injecting synthetic satellites into survey data and attempting to recover them with matched-filter searches in spatial and color–magnitude space. The resulting detection significance statistic, SIG, serves as a matched-filter signal-to-noise ratio: values above a given threshold indicate that a stellar overdensity is unlikely to be caused by noise or survey artifacts. A threshold of SIG $> 5.5$ is adopted as the baseline detection criterion, balancing the need to minimize false positives while reliably recovering true satellites (see Appendix D of \citealt{Drlica-Wagner2020}).

Detection probability is modeled as a function of absolute $V$-band magnitude ($M_V$), half-mass radius ($r_{h}$), and distance ($D$) of satellites. We base our correction on the 50\% completeness contours in the $(M_V, r_{h})$ plane, as a function of distance, provided in Figure 5 and 6 of \citet{Drlica-Wagner2020} (for DES and PS1) and Figure 5 of \citet{Tsiane2025} (for LSST). Where available, we use their analytical approximation of the 50\% detection limit, defined by the following expression:

\begin{equation}
\begin{split}
\log_{10}(r_{h,50}/{\rm pc}) =  &\ \frac{A_0(D/{\rm kpc})}{\left(M_V - M_{V,0}(D/{\rm kpc})\right)} + \\ & \log_{10}\left(r_{h,0}(D/{\rm kpc})\right),
\end{split}
\label{eq:detection_contour}
\end{equation}

\noindent where $r_{h,50}$ is the half-mass radius (in pc) corresponding to 50\% detectability, $M_V$ is the satellite's absolute $V$-band magnitude, and $D$ is the satellite's distance (in kpc). The distance-dependent parameters $A_0(D)$, $M_{V,0}(D)$, and $r_{h,0}(D)$ are calibrated for each survey and distance bin, with tabulated values provided in Table 5 of \citet{Drlica-Wagner2020} and Table 2 (Measured Star/Galaxy classification with the threshold of SIG $> 5.5$) of \citet{Tsiane2025}.

To translate this contour into a probabilistic detection model, we define a smooth transition function that assigns a detection probability based on the offset from the 50\% completeness line in the $\log_{10}(r_{h})$–$M_V$ plane. The probability is defined as:

\begin{equation}
P_{\mathrm{det}} = 
\left[ 1 + \exp\left( 
\frac{\log_{10}(r_{h}^{\mathrm{model}}) - \log_{10}(r_{h,50})}
{\Delta_{\mathrm{eff}}}
\right) \right]^{-1},
\label{eq:detection_probability}
\end{equation}

\noindent where $r_{h}^{\mathrm{model}}$ is the half-mass radius of the modeled galaxy,
$r_{h,50}$ is the half-mass radius defined by the 50\% completeness contour from Eq.~\eqref{eq:detection_contour},
$\Delta_{\mathrm{eff}}$ is an effective softening parameter that sets the sharpness of the transition. In our implementation, $\Delta_{\mathrm{eff}}$ is scaled inversely with $(M_{V,0} - M_V)$ to capture the sharper completeness boundary at brighter magnitudes using, 

\begin{equation}
\Delta_{\mathrm{eff}} = \frac{\Delta}{M_{V,0} - M_V},
\end{equation}

where $M_{V,0}$ is the survey-dependent reference magnitude from the completeness contour (Eq.~\ref{eq:detection_contour}). We adopt a value of $\Delta = 1$ dex (see Appendix 1 for a discussion on effects of varying this model and its parameter values). To maintain a stable transition when $|M_{V,0} - M_V| \rightarrow 0$, we impose a floor such that $\Delta_{\mathrm{eff}} \geq \Delta$. 

This function approaches 1 for galaxies below the 50\% contour (i.e., easily detectable) and falls off rapidly to 0 for galaxies above the contour (i.e., below the detection threshold). This probabilistic mask is applied to each model galaxy when constructing the predicted luminosity function to correct for systems likely to be detectable in a given survey. Further discussion of alternative implementations and different definitions of the detection probability can also be found in Appendix~\ref{AppA:Detection_prob}.\\

\section{Results}\label{result}

Using the model described in Section~\ref{method}, we predict the structural and photometric properties of satellite galaxies around MW-analogs, extending down to the hyper-faint regime. In this context, we define hyper-faint satellites as those with absolute $V$-band magnitudes $M_V > -3$ and half-stellar mass radii in the range $1 < R_{h}~[\mathrm{pc}] \lesssim 10$. These conditions are chosen to overlap with the region of the size–luminosity parameter space occupied by currently ambiguous stellar systems observed around the MW, as compiled in \citet{Pace2024}.

Figure~\ref{fig:size_luminosity} presents the predicted distribution of satellites in the $R_{h}$–$M_V$ plane. The colored regions represent the logarithm of the normalized probability density of satellites across all 100 MW-analogs. Darker regions indicate a higher likelihood of hosting satellites with those properties. The left panel (blue color map) shows results from our fiducial model, which includes H$_2$ cooling and UV background radiation effects. The right panel (green color map) shows results from a variant of the model in which H$_2$ cooling is excluded.

For comparison, observed MW satellites are shown with red circles, and ambiguous systems are marked in olive diamonds. At the bright end (lower $M_V$), both models reproduce the observed sizes of classical and ultra-faint satellites. Toward the faint end, both models predict a broader distribution in $R_{h}$ at fixed $M_V$. Notably, the predicted size–luminosity distributions in both models extend into the region occupied by ambiguous systems, suggesting that these low-luminosity, diffuse satellites may arise naturally within hierarchical models of galaxy formation.

\begin{figure*}
    \includegraphics[width=2.0\columnwidth]{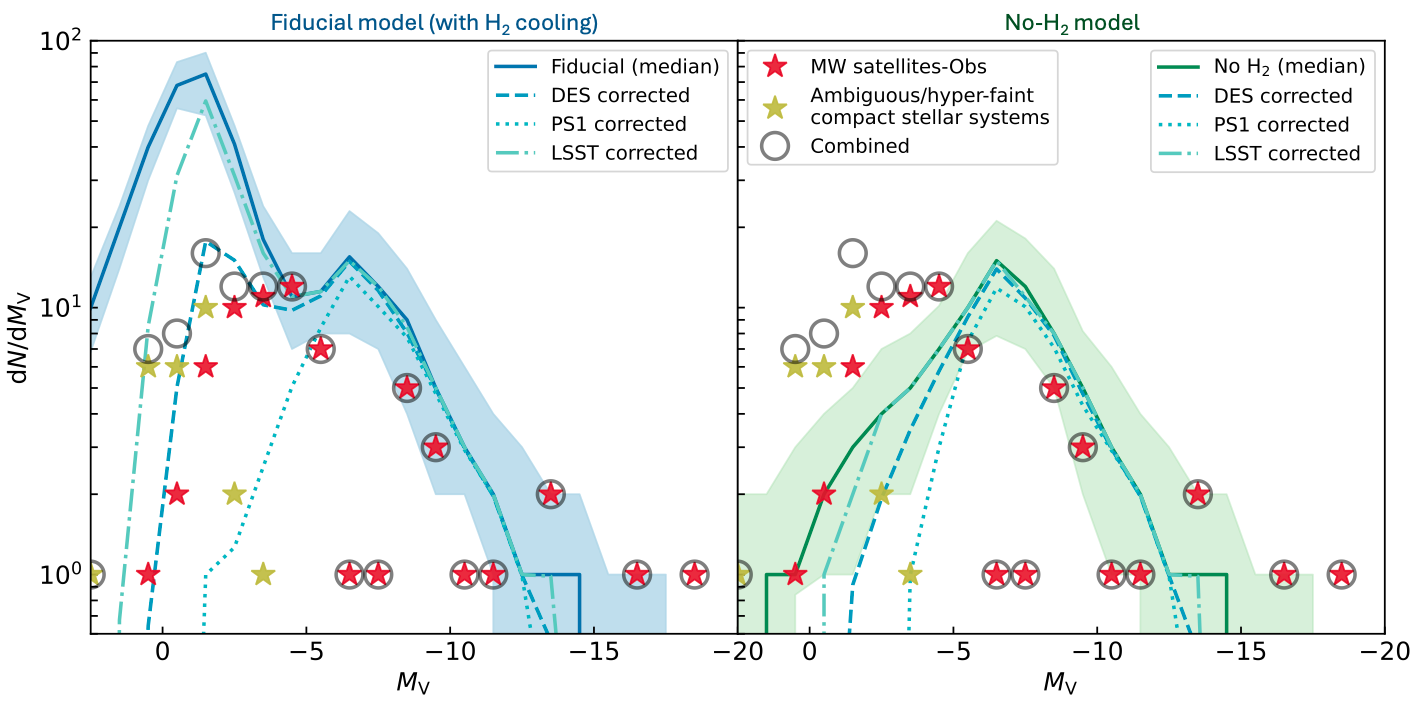}
    \caption{Luminosity function predictions for satellites of MW-analogs (line with the $1 \sigma$ halo-to-halo scatter). The left panel presents results from the fiducial model (including H$_2$ cooling), while the right panel shows results from the No-H$_2$ model (atomic cooling only). In both panels, corrections based on survey sensitivity have been shown for DES (dashed lines), PS1 (dotted lines), and LSST (dotted-dashed lines). For comparison, we include the observed luminosity function of confirmed MW satellites (red stars), ambiguous objects (olive stars), and the combined sample (black circles).}
    \label{fig:luminosity_function}
\end{figure*}

Comparing the two panels, we find that our fiducial model (incorporating H$_2$ cooling) produces a broader dispersion in size at fixed $M_V$ in the ultra-faint mass regime\footnote{``Ultra-faint satellites'' are classified as dwarf galaxies with absolute $V$-band magnitudes fainter than $M_V$ = $-7.7$ (L = 10$^5$ L$_{\odot}$, see \citealt{Simon2019}).}. This reflects the increased cooling efficiency and gas retention in small halos due to the contribution of molecular hydrogen. In {\sc Galacticus}, galaxy sizes are tied to the specific angular momentum content of the baryons (see Section 3.2.2 of \citealt{Ahvazi2024}), which depends on the host halo’s mass, spin, and the degree to which cooling and feedback preserve or expel angular momentum. With H$_2$ cooling included, lower-mass halos can form stars earlier, broadening the predicted size distribution at fixed luminosity (as will be shown in Figures 3 and 4, also see discussion in Section 3.1.1 of \citealt{Ahvazi2024}). The median trend in the size–luminosity relation under the fiducial model aligns more closely with that of the observed ultra-faint satellite population.

In addition to differences in the extent of the size–luminosity relation, our two models exhibit distinct locations for the peak probability of hosting satellites in this parameter space. This shift arises from the different thresholds for galaxy formation imposed by the presence or absence of H$_2$ cooling. Quantitatively, the number of hyper-faint satellites predicted by each model (measured within the region highlighted by the gray square in each panel) differs by nearly a factor of $\sim 7$, the fiducial model predicts roughly $\sim$ 40 such satellites per MW analog, compared to only about $\sim$ 5-6 in the No-H$_2$ model. To explore the broader satellite population more quantitatively, in Figure~\ref{fig:luminosity_function} we present the predicted satellite luminosity functions for each model. Similar to Figure~\ref{fig:size_luminosity}, the left panel corresponds to our fiducial model that includes both H$_2$ cooling and UV background radiation, while the right panel shows the results of the model without H$_2$ cooling.

In each panel, the solid line denotes the median predicted luminosity function across our sample of MW-analogs, and the shaded region represents the 1$\sigma$ halo-to-halo dispersion. The fiducial model (blue line and shaded region) predicts a larger number of ultra-faint and hyper-faint satellites compared to the model without H$_2$ cooling (green line and shaded region), which exhibits a decline in satellite counts at $M_V \gtrsim -6$.

We also include the observed luminosity function of the MW satellite population for comparison. Red markers represent confirmed satellites, olive markers indicate ambiguous systems, and the black circles show the combined set. Both models match the observed luminosity function reasonably well at the brighter (more massive) end. However, deviations appear in the ultra-faint and hyper-faint regimes.

It is important to emphasize that this faint end of the luminosity function is most susceptible to observational incompleteness due to the limited sensitivity of current surveys. Therefore, direct comparisons with observations without accounting for detectability limits may lead to biases. To address this, we apply survey-specific detection probabilities to our model predictions. These corrections are based on detection models from \cite{Drlica-Wagner2020} for the DES and PS1 surveys, and from \cite{Tsiane2025} for the LSST survey. The detection probability is a function of the satellite's V-band magnitude, half-mass radius, and  distance (see Section~\ref{sec:obs_incompleteness} for details). We apply these detection probabilities as weights when computing the luminosity functions. The resulting corrected luminosity functions for each model are shown in Figure~\ref{fig:luminosity_function}, with dashed, dotted, and dot-dashed lines corresponding to the DES, PS1, and LSST survey selection functions, respectively.

Interestingly, applying the DES detection probability to our fiducial model substantially reduces the predicted number of faint satellites and aligns the model closely with the currently observed satellite population. Furthermore, in the context of LSST's anticipated depth and coverage, our fiducial model predicts the discovery of dozens of new hyper-faint satellites (with $M_V >-3$), whereas the model without H$_2$ cooling predicts only a few. This again highlights the role of H$_2$ cooling in enabling galaxy formation at the lowest mass scales.

We also note that the corrected luminosity functions are somewhat sensitive to the exact form of the adopted detection probability function (Equation~\ref{eq:detection_probability} in Section~\ref{sec:obs_incompleteness}). In Appendix~\ref{AppA:Detection_prob}, we explore this sensitivity by varying the functional form and parameters, demonstrating the robustness of our main results within reasonable modeling uncertainties.

\begin{figure}
	\includegraphics[width=\columnwidth]{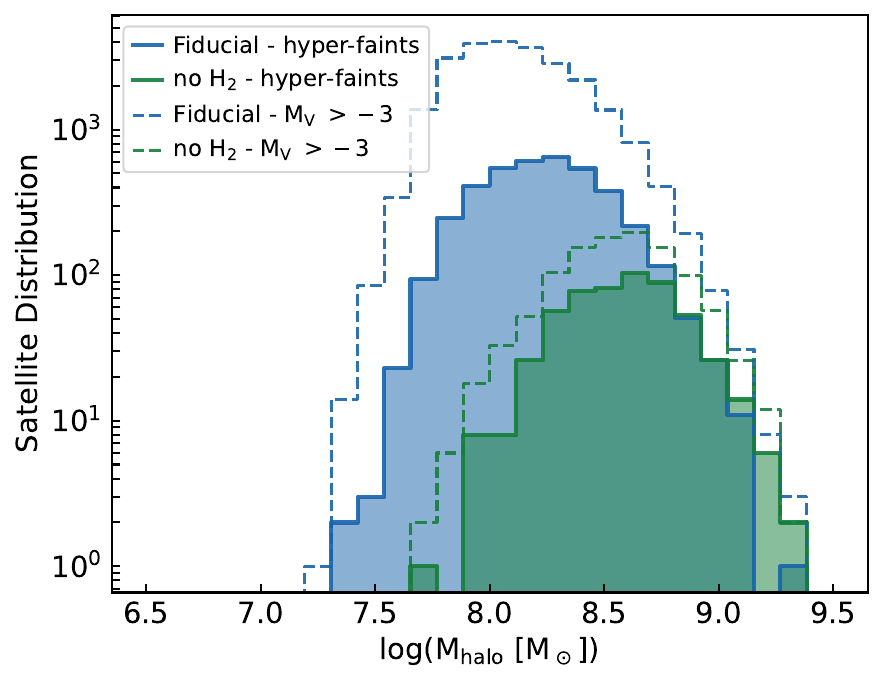}
    \caption{Comparison of peak halo masses for satellites with $M_V > -3$ (dashed lines) and for the subset classified as hyper-faint satellites (solid lines). Results from the fiducial model (including H$_2$ cooling) are shown in blue, while those from the No-H$_2$ model (atomic cooling only) are shown in green.}
    \label{fig:Mhalo}
\end{figure}

\subsection{Cooling Physics and Star Formation Histories of Hyper-Faint Satellites}

Beyond differences in satellite abundance, the properties of the hyper-faint satellite populations produced by each model vary significantly. Inclusion of H$_2$ cooling increases overall cooling efficiency, particularly at early times (before reionization suppresses H$_2$ formation), and during the epoch when low-mass halos are actively forming. As a result, star formation histories change such that surviving satellites at a given luminosity are hosted by halos with lower peak masses. Here, the peak halo mass refers to the maximum mass a subhalo attains over its assembly history, prior to any mass loss from tidal stripping.
Figure~\ref{fig:Mhalo} compares the peak halo masses of satellites with $M_V > -3$ in the fiducial (blue dashed) and No-H$_2$ (green dashed) models. The solid lines show the subset of systems classified as hyper-faint satellites according to our criteria ($M_V > -3$ {\bf and} $1 \lesssim R_h~[\mathrm{pc}] \lesssim 10$). As expected, hyper-faint satellites in the fiducial model tend to reside in lower-mass halos, reflecting the enhanced cooling efficiency enabled by molecular hydrogen.

\begin{figure}
	\includegraphics[width=\columnwidth]{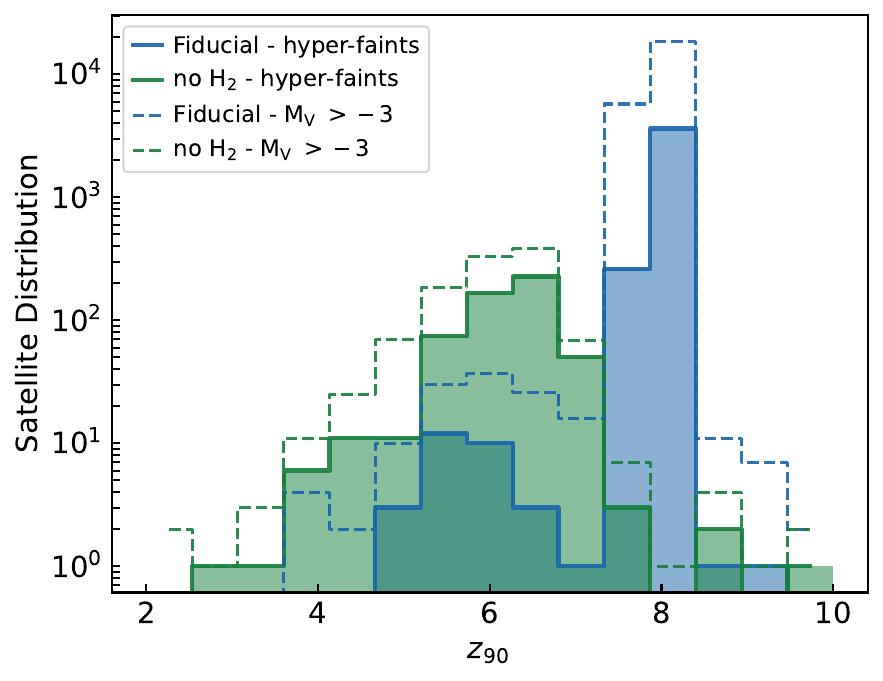}
    \caption{Comparison of star formation histories for satellites with $M_V > -3$ (dashed lines) and for the subset of hyper-faint satellites (solid lines). Results from the fiducial model (blue) and the No-H$2$ model (green) are shown. The comparison is based on $z_{90}$, defined as the redshift by which 90\% of a galaxy’s stellar mass has formed.}
    \label{fig:z90}
\end{figure}

In addition to halo masses, the timing of star formation also differs between the two models. Figure~\ref{fig:z90} compares the redshift of 90\% stellar mass formation ($z_{90}$)---that is, the redshift by which 90\% of a galaxy’s stellar mass has formed, analogous to the commonly used $\tau_{90}$. On average, hyper-faint satellites in the fiducial model exhibit earlier z$_{90}$ values, indicating older stellar populations and more rapid quenching histories. The z$_{90}$ distribution in the fiducial model is more sharply peaked at high redshift, suggesting
earlier quenching across the hyper-faint population when H$_2$ cooling is included. A smaller secondary peak appears at lower z, corresponding to a subset of hyper-faints residing in slightly more massive halos that can cool gas via atomic hydrogen and continue forming stars to later times, thereby resembling the delayed star formation seen in the No-H$_2$ model.

These results directly reflect the influence of the implemented cooling models. Once the universe is reionized two key processes impact star formation in low-mass halos: the increase in the UV background radiation and the heating of the intergalactic medium \citep[IGM;][]{Hambrick2011, Simpson2013, Noh2014, Pereira-Wilson2023}. In the fiducial model, the elevated background radiation rapidly dissociates H$_2$ molecules in low-mass halos, effectively shutting down star formation in the hyper-faint regime. Simultaneously, the heating of the IGM suppresses further gas accretion into these small halos, reinforcing the quenching process. Together, these effects lead to early and abrupt suppression of star formation in hyper-faint galaxies when H$_2$ cooling is included.

In contrast, when H$_2$ formation is excluded, hyper-faint satellites form in more massive halos that are less sensitive to photo-dissociation and thermal feedback. Although IGM heating still reduces gas accretion, its effect is mitigated because: (1) the host halos are more massive and thus better able to retain gas; and (2) there is a delay in the suppression of accretion, as it takes time for pressure waves to propagate and affect inflow. This results in a more gradual decline in star formation histories after reionization, as seen in the No-H$_2$ model. 

Current observational constraints on star‐formation histories in the hyper‐faint regime remain weak, owing to the small number of resolved stars in these systems and contamination effects. Deep, resolved photometry is required to reach the oldest main-sequence turnoff (oMSTO), which is challenging for systems with $M_V > -3$ because the oMSTO is faint and many objects have only few stars detected around that region. For example, \citet{Durbin2025} analyze resolved star formation histories for 36 ultra‐faint dwarf galaxies within $\lesssim 400$ kpc using HST imaging that reaches oMSTO. From their full sample, a subset of $\sim 20$ systems have $M_V > -3$, comparable in luminosity to our hyper‐faint sample. Within that subset, they report a large spread in $\tau_{90}$ (the range of $\sim 6.2 - 13.35$ Gyr lookback time by which 90\% of stellar mass has formed). However, they caution that for many of these faint objects their $\tau_{90}$ estimates carry large uncertainties. In fact, if one restricts to those with $\tau_{90}$ error bars smaller than $\sim 3$ Gyr, only about eight objects remain; these tend to be biased toward later quenching times (i.e. lower $z_{90}$), because earlier quenching (very high $z_{90}$) is harder to detect reliably given their data and modeling limitations.

In our fiducial model, which includes H$_2$ cooling, many hyper‐faint satellites are predicted to form the majority of their stars very early (high $z_{90}$), followed by rapid quenching near the epoch of reionization. Because our model assumes a spatially uniform UV background (meaning that all halos experience reionization suppression simultaneously), it does not capture the effects of patchy reionization, where gas suppression could occur at different times depending on local ionization fronts or environmental shielding. Although {\sc Galacticus} includes a simplified clumping factor prescription (see section 2.3 in \citealt{Ahvazi2024} for details and Section~\ref{sec:modeling_caveat} of this work for additional modeling assumptions affecting star formation histories), a more realistic treatment of spatially inhomogeneous reionization could alter the predicted $z_{90}$ distribution. Consequently, comparing our predicted $z_{90}$ values with measurements from samples such as \citet{Durbin2025} offers a way to constrain not only the physics of H$_2$ cooling but also the timing and potential spatial variation of reionization (provided that sufficiently precise star formation histories can be measured for larger numbers of low‐mass systems).

\subsection{Kinematic Signatures: Line-of-Sight Velocity Dispersion Predictions}

\begin{figure}	\includegraphics[width=\columnwidth]{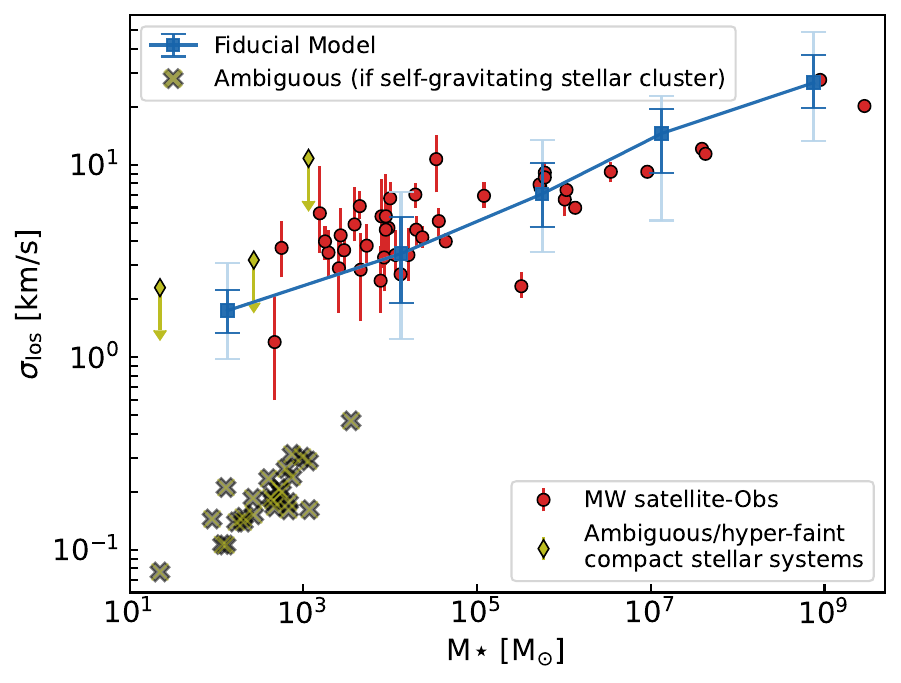}
    \caption{Line-of-sight velocity dispersion predictions (measured at r$\rm_{h}$) from our fiducial model, shown as the blue line, as a function of satellite stellar mass. For comparison, we include observed velocity dispersions of confirmed MW satellites (red markers) and the limited available measurements for ambiguous systems (olive markers). We also show the estimated velocity dispersions these ambiguous objects would have if they were purely self-gravitating stellar clusters (olive crosses with black edges).}
    \label{fig:vel_dispersion}
\end{figure}

Having examined how the properties of hyper-faint satellites vary across our cooling models, we now turn to their kinematic signatures. Specifically, we focus on predictions from our fiducial model for line-of-sight velocity dispersion ($\sigma_{\rm los}$) measurements to help distinguish between bona fide satellite galaxies and ambiguous objects that may instead be stellar clusters devoid of dark matter. 

In {\sc Galacticus}, line-of-sight velocity dispersions are derived by solving the Jeans equation (assuming isotropic velocities) in the combined potential of DM halo and galaxy. The stellar density profile $\rho_\star(r)$ and total enclosed mass $M(r)$ are used to compute the radial dispersion $\sigma_r(r)$, which is then projected along the line of sight and weighted by the stellar distribution:
\begin{equation}
\sigma_{\rm los}^2(R) = \frac{\int_{-\infty}^{\infty} \rho_\star(r)  \sigma_r^2(r)  dl}{\int_{-\infty}^{\infty} \rho_\star(r)  dl}, \quad r = \sqrt{R^2 + l^2}.
\end{equation}
This approach differs from the \citet{Wolf2010} mass estimator, which relates $\sigma_{\rm los}$ directly to the mass within the half-light radius. By solving the Jeans equation and averaging over radii, the {\sc Galacticus} method provides a more general prediction for $\sigma_{\rm los}$ that directly reflects the depth of the gravitational potential.
Figure~\ref{fig:vel_dispersion} presents predictions for $\sigma_{\rm los}$ from our fiducial model (blue curve), with error bars indicating the 1$~\sigma$ and 2$~\sigma$ confidence intervals. Our model predictions align with observed velocity dispersions of MW satellites, shown as red markers on this figure.

Our predictions extend into the regime of hyper-faint systems where there is no robust measurement of velocity dispersions. For most of these systems, current data only provide upper limits, shown as olive-colored markers in Figure~\ref{fig:vel_dispersion}. These few available measurements appear broadly consistent with the predictions of our fiducial model, which yields $\sigma_{\rm los} \sim 1–3$ km/s, though the uncertainties remain large.

To explore whether these ambiguous objects could be consistent with self-gravitating stellar systems that are devoid of dark matter, we estimate their expected line-of-sight velocity dispersions assuming purely stellar profiles. Following the methodology outlined in \cite{Errani2024}, we model the stellar distribution as a spherically symmetric exponential profile:
\begin{equation}
    \rho_*(r) = \rho_0 \, e^{-r / r_*},
\end{equation}

\noindent where $\rho_0$ is the central density and $r_*$ is the 3D scale radius of the stellar component. The corresponding projected (2D) half-light radius R$_h$ is related to the scale radius via $R_h \approx 2.03 \, r_*$ \cite[see section 3.3.2]{Errani2024}. Assuming the system is in virial equilibrium and fully self-gravitating, the luminosity-weighted, line-of-sight velocity dispersion can be expressed as:

\begin{equation}
    \langle \sigma_{\text{los}}^2 \rangle = \frac{5}{96} \, \frac{G M_*}{r_*},
\end{equation}

\noindent where  M$_*$ is the total stellar mass and G is the gravitational constant. We use this expression to calculate $\sigma_{los}$ for each ambiguous object using their observed stellar mass and projected half-light radius. These estimates are shown as gray circles overlaid with olive ``×'' markers in Figure~\ref{fig:vel_dispersion}.
The results show that for objects with similar stellar mass, the self-gravitating, dark matter–free scenario predicts $\sigma_{\rm los}$ values that are approximately an order of magnitude lower than those predicted by our fiducial model, which includes the contribution of dark matter. This discrepancy offers a promising avenue to assess the nature of these systems. Improved measurements of $\sigma_\mathrm{los}$ in the hyper-faint regime (objects currently classified as ambiguous) could therefore provide a powerful diagnostic for distinguishing between dark matter–dominated dwarf galaxies and star clusters.

\section{Discussion}\label{discussion}

In this section, we place our results in a broader context by considering both the limitations of our modeling framework and the observational uncertainties that affect the classification of ambiguous systems.

\subsection{Interpreting Results in the Context of Model Assumptions}\label{sec:modeling_caveat}

The No-H$_2$ cooling model provides a useful baseline for isolating the role of molecular hydrogen cooling, but it does not include the effects of photoheating on atomic gas. This omission may be significant, especially when studying the impact of reionization on the star formation histories of hyper-faint galaxies. Photoheating can rapidly suppress star formation in low-mass halos by counteracting the cooling which otherwise allows gas to lose pressure support and flow into the center of the halo to form stars. If the photoheating rate exceeds the atomic cooling rate, gas in these halos cannot cool efficiently and star formation is shut down shortly after reionization. This effect can quench star formation more rapidly than the gradual reduction in gas accretion due to IGM heating, and its exclusion could lead to an overestimate in the number of hyper-faint systems predicted by this model.  
Incorporating photoheating effects into the No-H$_2$ model would likely further suppress star formation in low-mass halos. Already our No-H$_2$ cooling model produces a relatively small sample of hyper-faint satellites—approximately 550 in total—compared to about 3,900 in the fiducial model. This corresponds to an average of $\sim 5$ hyper-faint satellites per MW–analog merger tree, versus $\sim 40$ per tree in the fiducial model. The addition of photoheating could potentially reduce the average number of hyper-faint satellites from $\sim 5$ to nearly zero.

When interpreting our results, it is important to keep in mind that stellar tidal heating is not explicitly modeled in the current framework, which instead focuses on the dominant dark matter stripping processes in both models presented. Prior work has demonstrated that tidal interactions with the host halo/central galaxy can significantly influence the structural and dynamical properties of satellites, particularly through tidal stripping. Given that the satellite systems under study have extremely low stellar masses and low stellar-to-dark matter mass ratios, we expect tidal stripping to predominantly affect the dark matter halos—an effect that is included in our current modeling framework \citep{Du2024}. However, some level of stellar mass loss is also expected, especially in systems that have experienced extensive tidal stripping.

Results by \cite{Penarrubia2008} and \cite{Errani2024} provide a useful benchmark for assessing when stellar components begin to exhibit signs of tidal disruption. Their study shows that noticeable changes in the stellar distribution typically occur only after extreme mass loss, with more than 99\% of the original subhalo mass needing to be stripped (i.e., when the remnant mass fraction drops below $10^{-2}$). 
In our analysis, most surviving satellites remain above this threshold. For the hyper-faint satellite population, the median bound mass fractions are $0.28^{+0.55}_{-0.13}$ in our fiducial model and $0.22^{+0.46}_{-0.08}$ in the No-H$_2$ model. These values indicate that while many systems have experienced significant dark matter loss, the majority have not yet reached the most extreme stripping regime. 
That said, the absence of an explicit prescription for stellar tidal heating and stripping in our current modeling remains an important limitation, and incorporating a more sophisticated treatment of stellar tidal heating in future work will be essential for accurately modeling the most heavily stripped satellites and for making robust predictions about their stellar morphology and detectability.

A final caveat relates to our choice of host halos. The MW analogs studied here were selected to have present-day ($z=0$) halo masses similar to that of the MW, but they do not necessarily share its detailed formation history, including the presence of a Large Magellanic Cloud (LMC) analog or a relatively quiescent merger history prior to the LMC’s infall. Both factors can affect the surviving satellite population \citep{Buch2024}. Within our sample of 100 MW-mass halos, roughly 16 contain an LMC-mass analog as their brightest satellite within 300 kpc of the central galaxy\footnote{Here, an LMC analog is defined loosely as any satellite brighter than M$_V = -18$ within 300 kpc of the central galaxy. This simple stellar-mass-based selection does not account for additional criteria such as orbital history, infall time, or the presence of an SMC companion.}. While this provides a useful range of possible environments, it does not capture the full merger history of the LMC analog itself, which could further shape the properties of the faintest satellites.

\subsection{Star Cluster or Galaxy}

Our analysis indicates that hyper-faint satellites are expected to have small line-of-sight velocity dispersions, on the order of $\sigma_{\rm los} \sim 1$–3 km/s (Figure~\ref{fig:vel_dispersion}). By contrast, purely self-gravitating stellar systems of comparable luminosity would yield values nearly an order of magnitude lower. This difference makes velocity dispersion a powerful diagnostic for distinguishing dark matter–dominated dwarf galaxies from star clusters.

Discriminating between the two scenarios, however, requires precise measurements of $\sigma_{\rm los}$ in the hyper-faint regime. Obtaining such measurements remains extremely challenging. Contamination from foreground MW stars and the presence of unresolved binary stars within the observed sample can significantly bias estimates of the line-of-sight velocity dispersion (see \citealt{Simon2019} for a detailed discussion). In particular, binary orbital motions introduce a systematic inflation of the measured dispersion that becomes increasingly problematic at low intrinsic values \citep[e.g.][]{McConnachie2010, Minor2010, Gration2025}. Long-period binaries ($P \gtrsim$ a few years) alone can generate apparent dispersions of order $\sim 1$ km/s in purely self-gravitating stellar systems with no dark matter, provided the binary fraction is sufficiently high. Using simulated galaxies from the AURIGA suite, \citet{Wang2023} demonstrate that binaries can artificially elevate the inferred velocity dispersion (and the dynamical mass), with the impact growing rapidly for $\sigma_{\rm los} \lesssim 3$ km/s. For binary fractions as high as 70\%, they find that the dynamical mass within the half-light radius can be overestimated by 10–15\% even after two-epoch observations (separated by 1 year) designed to remove short-period binaries, and by up to 60\% in extreme cases at $\sigma_{\rm los} \sim 1$ km/s. Such inflation can mimic the dynamical signal expected from dark matter, increasing the risk of misclassifying some ambiguous objects as dwarf galaxies. These challenges make it particularly difficult to obtain robust kinematic measurements in the hyper-faint regime and highlight the importance of  binary correction strategies in future observational efforts.

Recent work has added further complexity to this classification problem. Independent studies have proposed alternative explanations for Ursa Major III/UNIONS 1 (UMa3/U1), a particularly intriguing “ambiguous” system lying near the boundary between galaxies and star clusters\footnote{Based on recent analysis by \citet{Cerny2025}, UMa3/U1 is now considered a likely star cluster.}. \citet{Rostami2025} use $N$-body simulations to argue that UMa3/U1’s high dynamical mass-to-light ratio could arise without dark matter if it is a star cluster hosting a centrally concentrated black hole subsystem that inflates its velocity dispersion. Similarly, \citet{Devlin2025} employ collisional $N$-body simulations (including stellar evolution and the MW tidal field) and show that with inclusion of primordial binaries they can also produce elevated velocity dispersions consistent with the observations. In contrast, \citet{Errani2024b} demonstrate that a dark-matter–dominated “microgalaxy” model naturally explains UMa3/U1’s compact size, elevated dispersion, and long-term survival. These results underscore the need for complementary diagnostics beyond kinematics. For example, \citet{Devlin2025} highlight that mass segregation, which is common in globular clusters with ages exceeding their relaxation times (e.g., \citealt{Baumgardt2022}), naturally leads to preferential loss of low-mass stars in the cluster outskirts under tidal stripping, resulting in a top-heavy present-day stellar mass function. In contrast, dark-matter–dominated galaxies are expected to have minimal mass segregation since the stellar component remains dynamically well-mixed due to the long relaxation times of galaxies. As a result, their present-day mass function should closely resemble their initial mass function. This “mass function test” has been proposed as a diagnostic for distinguishing between dark matter–dominated ultra-faint galaxies and self-gravitating star clusters (granted deeper photometric data are available from “ambiguous” objects). Our sample of hyper-faint satellites are highly dark matter dominated (even after losing on average $\sim 80\%$ of their halo mass) which should help them retain their stellar populations and preserve an IMF-like present-day mass function. This makes them excellent targets for such tests and provides a pathway to classify currently ambiguous systems.

Other observational methods can aid in differentiating low-mass satellite galaxies from stellar clusters. One promising avenue is the study of stellar metallicity distributions. Satellite galaxies are generally expected to host more metal-poor and chemically diverse stellar populations compared to globular clusters of similar luminosity. This is due to their extended star formation histories and the retention of supernova ejecta in deeper gravitational potential wells. High-resolution spectroscopy to measure metallicity spreads and elemental abundance patterns can thus provide critical clues about the formation history and dark matter content of these faint systems. 

Importantly, the converse is not necessarily true: the absence of a measurable metallicity spread does not rule out a galaxy classification, particularly for hyper-faint systems in which a single, brief star-formation episode can dominate the stellar population and mask any potential spread \citep{Webster2016}. 

Building on this, recent hydrodynamical zoom-in simulations have predicted the existence of a population of globular-cluster–like dwarfs (GCDs) that naturally occupy the parameter space between classical globular clusters and ultra-faint dwarf galaxies \citep{Taylor2025}. These objects form in low-mass dark matter halos at high redshift, undergo a single short star-formation episode, and self-quench, resulting in very low metallicities, narrow but measurable metallicity spreads ($\sim 0.1–0.3$ dex), and moderate dynamical mass-to-light ratios ($\approx$ tens M$_\odot/$L$_\odot$). Their structural properties and internal kinematics closely overlap with several known “ambiguous” systems, including Reticulum II, Boötes V, and Draco II. Our hyper-faint sample, which resides in dark-matter–dominated halos and shows evidence for early quenching, may therefore overlap with the predicted GCD population. Detailed measurements of metallicity spreads, dynamical mass-to-light ratios, and age distributions of “ambiguous” objects will be critical for determining whether these systems are indeed analogues of GCDs, which would have important implications for the survival of the smallest dark matter halos and for constraints on models with suppressed small-scale structure.

Our fiducial predictions are based on the standard cold dark matter (CDM) framework and do not incorporate alternative dark matter models such as warm dark matter (WDM) or self-interacting dark matter (SIDM). These models can lead to different predictions for both the abundance and internal structure of satellite galaxies. For example, WDM models suppress the formation of low-mass halos due to the cutoff in the linear matter power spectrum, in extreme cases potentially eliminating hyper-faint galaxies altogether (regardless of the details of baryonic physics). SIDM models, can lead to the formation of central density cores in low-mass halos, modifying the mass distribution and potentially reducing the predicted velocity dispersions and enhancing tidal disruption, which lowers the surviving  satellite abundance. Because the existence (or absence) of hyper-faint satellites is highly sensitive to the initial power spectrum and the nature of dark matter interactions, identifying these systems carries strong implications for testing alternative dark matter models. Testing these alternative scenarios requires more accurate observational constraints, especially on key properties such as stellar velocity dispersion and structural parameters, which can help distinguish between different dark matter models.

\section{Summary and Conclusions}\label{conclusions}

In this work, we explored the formation and properties of hyper-faint satellite galaxies (defined as those with M$_V > -3$ and half-mass radii between 1--10 pc) in MW-like halos using the semi-analytic galaxy formation model {\sc Galacticus}. To isolate the effects of H$_2$ cooling coupled with UV background radiation on the formation and evolution of these faint satellite population, we compare two models variants: a fiducial model that includes both H$_2$ cooling and UV background radiation, and a No-H$_2$ model that omits these processes.

The fiducial model successfully reproduces the structural and photometric properties of known MW satellites, as well as their overall abundances once observational incompleteness is taken into account. Its predictions in size–luminosity space extend into the hyper-faint regime, where it predicts a substantial population of faint systems. Notably, this region of parameter space overlaps with that occupied by several discovered “ambiguous” systems, suggesting that such low-luminosity satellites may arise naturally within hierarchical galaxy formation scenarios.

Our results highlight the critical role of H$_2$ cooling in setting the threshold for galaxy formation and enabling star formation at the smallest mass scales, by allowing lower-mass halos to host galaxies (see Figure~\ref{fig:Mhalo}). While both the fiducial and No-H$_2$ models are capable of forming hyper-faint satellites, the inclusion of H$_2$ cooling leads to a larger number of faint systems (Figure~\ref{fig:luminosity_function}), broader size–luminosity distributions (Figure~\ref{fig:size_luminosity}), and earlier quenching of star formation in the hyper-faint regime (Figure~\ref{fig:z90}). The fiducial model predicts an average of $\sim$40 hyper-faint satellites per MW analog, compared to only $\sim$5–6 in the No-H$_2$ model. Many of these systems would lie below current survey detection limits but are likely within reach of upcoming surveys such as LSST, Roman, and Euclid.

We further predict that hyper-faint galaxies in the fiducial model have line-of-sight velocity dispersions of $\sigma_{\rm los} \sim 1–3$ km/s, nearly an order of magnitude higher than expected for purely self-gravitating stellar clusters of comparable stellar mass (Figure~\ref{fig:vel_dispersion}). This contrast provides a key diagnostic for identifying whether newly discovered ultra-faint systems are dark matter–dominated galaxies or star clusters. However, achieving the precision required to measure such small dispersions, especially in systems with few observable stars, remains a major observational challenge.

\section*{Acknowledgements}

We gratefully acknowledge the contributions of several individuals who supported this research. We thank Anatoly Klypin, Isabel Santos-Santos, and Mark Whittle for insightful discussions that helped shape this work. We are also grateful to Alex Drlica-Wagner and Kabelo Tsiane for sharing the data from \citet{Drlica-Wagner2020, Tsiane2025}. We acknowledge useful discussions during Galaxy Evolution and Cosmology (GECO) lunches at the University of Virginia. The computations presented here were conducted using the High Performance Computing resources at Carnegie Institution for Science and the University of Virginia.

\section*{Data Availability}

The SAM model utilized in this project is openly accessible and can be accessed at the following link:
\href{https://github.com/galacticusorg/galacticus/ee197e314ac28f1eca3b6dcc47c5f88682bbddb5}{https://github.com/galacticusorg/galacticus/}. Researchers and interested parties can freely explore and utilize the SAM model to replicate and build upon the findings presented in this study.


\bibliographystyle{mnras}
\bibliography{manuscript}



\appendix
\section{Appendix A: Comparison of Parametric Detection Probability Models} \label{AppA:Detection_prob}

\begin{figure*}
    \center
    \includegraphics[width=\columnwidth]{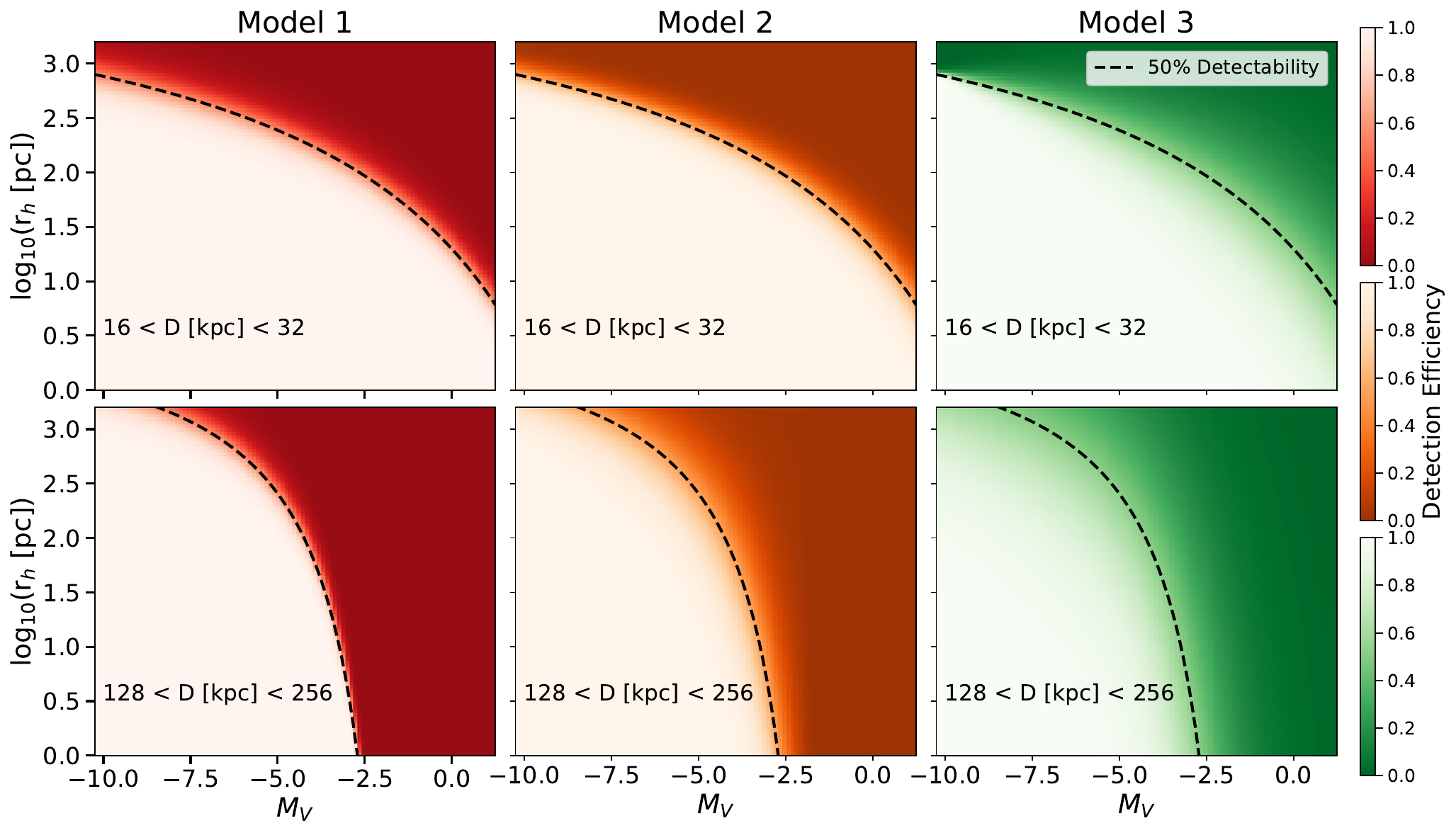}
    \caption{Comparison of three models for predicting the detection probability of MW satellites in size–luminosity space (r$_h$–M$_V$). Each column shows a different model (Model 1: red, Model 2: orange, Model 3: green). As an example, we apply the models to the DES survey, using the 50\% completeness contours (dashed black lines in each panel) as a reference. The upper and lower panels correspond to two representative distance bins.}
    \label{fig:detection_prob}
\end{figure*}

To account for survey incompleteness, we implemented several parametric forms for the detection probability function, each designed to translate the 50\% completeness contours of \citet{Drlica-Wagner2020} and \citet{Tsiane2025} into a smooth probability mask. Here we compare three variants, which we refer to as Model 1, Model 2, and Model 3. All models adopt the same underlying completeness contour,

\begin{equation}
\log_{10}(r_{h,50}/{\rm pc}) = \frac{A_0(D)}{M_V - M_{V,0}(D)} + \log_{10}\left(r_{h,0}(D)\right),
\end{equation}

\noindent where $r_{h,50}$ is the half-mass radius at which detection efficiency drops to 50\%, $M_V$ is the absolute $V$-band magnitude, and $D$ is the distance. The parameters $A_0$, $M_{V,0}$, and $r_{h,0}$ are survey- and distance-dependent calibration constants. The completeness contour is then converted into a detection probability through a sigmoid function,

\begin{equation}
P_{\mathrm{det}} = 
\left[ 1 + \exp\left( \frac{X}{\Delta_{\mathrm{eff}}} \right) \right]^{-1},
\label{eq:detection_probability}
\end{equation}

\noindent where $X$ quantifies the offset of a model galaxy from the 50\% contour and $\Delta_{\rm eff}$ sets the steepness of the transition. The three implementations differ primarily in how $X$ and $\Delta_{\rm eff}$ are defined. To illustrate these differences, Figure~\ref{fig:detection_prob} shows the detection probabilities for each model applied using the DES 50\% completeness contours, with each model displayed in a separate column. The upper and lower panels correspond to two representative distance bins. A detailed description of each implementation is provided below.

\subsection{A.1 Model 1: Fixed–width sigmoid in $\log r_h$ space}

In Model 1, the offset is defined simply as the vertical distance in $\log_{10}(r_h)$ from the completeness contour,

\begin{equation}
X = \log_{10}(r_h^{\rm model}) - \log_{10}(r_{h,50}),
\end{equation}

with a fixed transition width $\Delta_{\rm eff} = \Delta$ (default $\Delta = 0.1$ dex). This form produces a symmetric, magnitude-independent smoothing around the 50\% line, as shown in the left column of Figure~\ref{fig:detection_prob}. It is simple and efficient but does not capture the fact that completeness transitions are sharper for brighter systems and smoother for fainter systems (see figures 5 in \citealt{Drlica-Wagner2020} and \citealt{Tsiane2025}). 

\subsection{A.2 Model 2: Magnitude–dependent scaling of the transition width}

Model 2 modifies the transition width to scale inversely with the magnitude difference from the reference value,

\begin{equation}
\Delta_{\rm eff} =
\begin{cases}
\frac{\Delta}{M_{V,0} - M_V} & |M_{V,0} - M_V| > 1, \\
\Delta & \text{otherwise}.
\end{cases}
\end{equation}

This scaling sharpens the probability cutoff for bright satellites while maintaining a floor at $\Delta = 1$ for fainter systems to prevent the transition from becoming unrealistically large near the singularity of the denominator. The resulting behavior is illustrated in the middle column of Figure~\ref{fig:detection_prob}. Compared to Model 1, this functional form provides a more realistic representation of how detection efficiency declines toward fainter magnitudes.

\subsection{A.3 Model 3: Two–dimensional distance metric to the completeness contour}

Model 3 generalizes the offset definition by computing the minimum Euclidean distance in the $(M_V, \log r_h)$ plane between a model satellite and the completeness contour. That is,
\begin{equation}
X = \pm \sqrt{(M_V - M_V^\star)^2 + \left(\log_{10} r_h - \log_{10} r_{h,50}(M_V^\star)\right)^2},
\end{equation}
\noindent where $(M_V^\star, r_{h,50}(M_V^\star))$ is the point on the contour closest to the model galaxy. The sign is chosen such that $X>0$ corresponds to galaxies above the detection threshold. The transition width is further allowed to vary linearly with $M_{V,0} - M_V$,
\begin{equation}
\Delta_{\rm eff} = \alpha (M_{V,0} - M_V) + \Delta,
\end{equation}
with $\Delta = 0.75$ and $\alpha = -0.04 $ introducing sharper transitions for brighter galaxies. This construction provides the most flexible mapping, capturing both horizontal and vertical offsets from the detection boundary. 

\subsection{A.4 Comparison}

\begin{figure}
    \center
	\includegraphics[width=0.6\columnwidth]{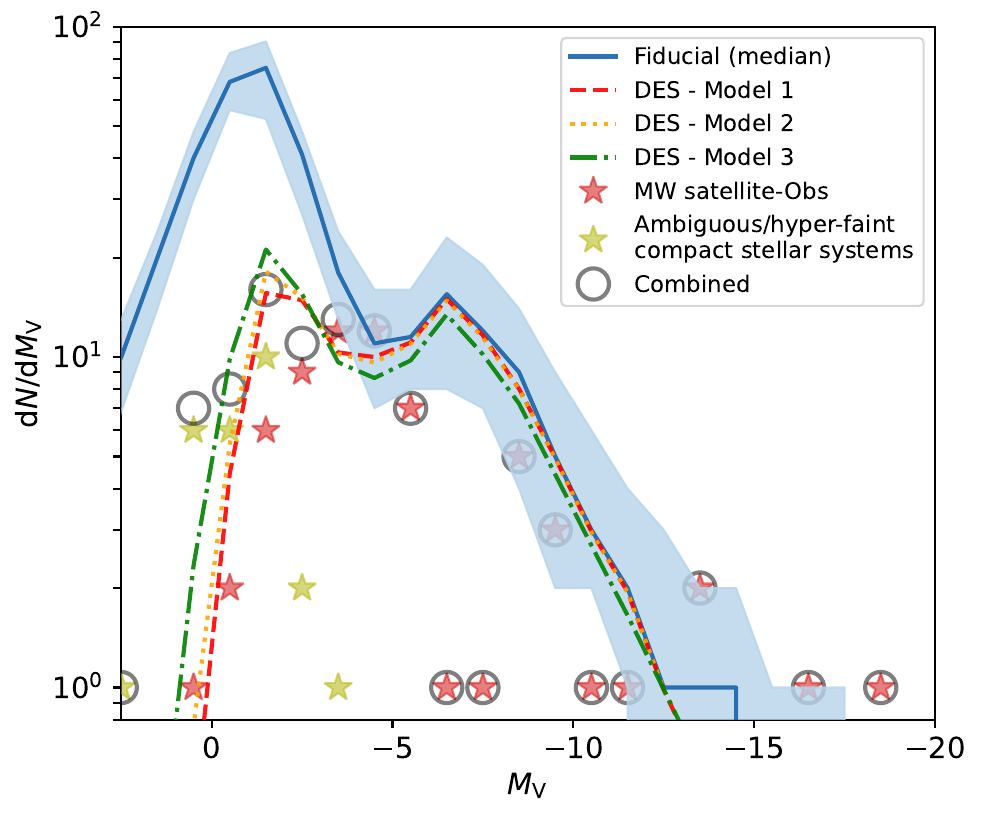}
    \caption{Comparison of the effects of the three detection-efficiency models on the predicted satellite luminosity functions. Each model is shown with a different line style and color (Model 1: dashed red, Model 2: dotted orange, Model 3: dash-dotted green). As an example, we apply the models using the 50\% completeness contours of the DES survey to the satellites predicted by our fiducial model (solid blue line, with shaded region showing the 1$\sigma$ halo-to-halo dispersion). For reference, we also include the observed MW satellites (red markers), ambiguous objects (olive markers), and their combined population (gray markers).}
    \label{fig:lum_function_DES}
\end{figure}

We applied all three detection-probability models to the 50\% completeness contours of the DES survey. Figure~\ref{fig:lum_function_DES} shows the resulting satellite luminosity functions after applying the detection efficiencies, with each model plotted in the same colors as in Figure~\ref{fig:detection_prob}. Overall, the three models yield broadly consistent luminosity functions, differences are most noticeable near the faint limit, where the functional form of the transition plays a larger role, but even there the predictions remain consistent within the expected uncertainties.
Given this agreement, we adopt Model 2 as our fiducial choice. It provides a balance between simplicity and realism, going beyond the overly simplistic fixed-width approach of Model 1, while avoiding the additional complexity of Model 3, by adjusting only the scaling of the transition width. This model is therefore applied consistently across all surveys (DES, PS1, and LSST) when accounting for detection efficiencies in our luminosity function calculations.\\
\end{document}